\documentclass[letterpaper, 10pt, conference]{ieeeconf}
\usepackage{newtxtext,newtxmath}
\usepackage[scaled=0.92]{helvet}
\usepackage{courier}
\usepackage{amsmath}
\usepackage{graphicx}
\usepackage{tikz}
\usepackage{bondgraphs}
\usepackage{url}
\usetikzlibrary{external}

\usepackage{float}
\usepackage{outlines}
\usepackage[font=footnotesize]{caption}
\usepackage[font=footnotesize]{subcaption}

\usepackage[dvipsnames]{xcolor}
\usepackage{circuitikz}
\usetikzlibrary{arrows.meta, decorations.markings, positioning, calc, shapes.geometric}

\usepackage{pgfplots}
\pgfplotsset{compat=1.18}
\usepgfplotslibrary{fillbetween}

\definecolor{mplC0}{HTML}{0072B2} 
\definecolor{mplC1}{HTML}{D55E00} 
\definecolor{mplC4}{HTML}{009E73} 
\definecolor{mplC5}{HTML}{CC79A7} 
\definecolor{mplC6}{HTML}{E69F00} 
\definecolor{mplC7}{HTML}{56B4E9} 
\definecolor{mplC8}{HTML}{000000} 

\usepackage[backend=biber,
  style=ieee,
  minnames=1,
  maxcitenames=2, maxbibnames=6,
  doi=false,
  isbn=false,
  url=false,
  eprint=false
]{biblatex}

\DeclareFieldFormat{journaltitle}{\mkbibemph{#1}}
\DeclareFieldFormat{booktitle}{\mkbibemph{#1}}

\DeclareFieldFormat[article,inproceedings,patent]{titlecase}{\MakeSentenceCase*{#1}}

\DeclareSourcemap{
  \maps[datatype=bibtex]{
    \map{
      \step[fieldset=language, null]
      \step[fieldset=abstract, null]
      \step[fieldset=keywords, null]
      \step[fieldset=file, null]
      \step[fieldset=urldate, null]
      \step[fieldset=note, null]
    }
    \map[overwrite]{
      \step[fieldsource=journal, match=\regexp{(?i)\bieee\b}, replace={IEEE}]
      \step[fieldsource=journaltitle, match=\regexp{(?i)\bieee\b}, replace={IEEE}]
      \step[fieldsource=booktitle, match=\regexp{(?i)\bieee\b}, replace={IEEE}]
      \step[fieldsource=journal, match=\regexp{(?i)\biet\b}, replace={IET}]
      \step[fieldsource=journaltitle, match=\regexp{(?i)\biet\b}, replace={IET}]
    }
    \map[overwrite]{
      \step[fieldsource=title, match=\regexp{\bEI\b}, replace={{EI}}]
      \step[fieldsource=title, match=\regexp{\bFONS\b}, replace={{FONS}}]
      \step[fieldsource=title, match=\regexp{\bHodgkin-Huxley\b}, replace={{Hodgkin-Huxley}}]
      \step[fieldsource=title, match=\regexp{\bHodgkin--Huxley\b}, replace={{Hodgkin--Huxley}}]
      \step[fieldsource=title, match=\regexp{\bport-Hamiltonian\b}, replace={{port-Hamiltonian}}]
    }
  }
}
\IEEEoverridecommandlockouts                                     
\newcommand{\emem}{e_{\mathrm{m}}}
\newcommand{\es}{e_{\mathrm{s}}}
\newcommand{\Cm}{C_{\mathrm{m}}}
\newcommand{\Cs}{C_{\mathrm{s}}}
\newcommand{\qm}{q_{\mathrm{m}}}
\newcommand{\Vm}{V_{\mathrm{m}}}

\newcommand{\gL}{g_{\mathrm{L}}}
\newcommand{\gLbar}{\bar{g}_{\mathrm{L}}}
\newcommand{\RL}{R_{\mathrm{L}}}
\newcommand{\EL}{E_{\mathrm{L}}}
\newcommand{\fL}{f_{\mathrm{L}}}

\newcommand{\gE}{g_{\mathrm{E}}}
\newcommand{\gEbar}{\bar{g}_{\mathrm{E}}}
\newcommand{\EE}{E_{\mathrm{E}}}
\newcommand{\fE}{f_{\mathrm{E}}}
\newcommand{\RE}{R_{\mathrm{E}}}
\newcommand{\kE}{\kappa_{\mathrm{E}}}
\newcommand{\eEth}{e_{\mathrm{E}}^{\mathrm{th}}}

\newcommand{\gI}{g_{\mathrm{I}}}
\newcommand{\gIbar}{\bar{g}_{\mathrm{I}}}
\newcommand{\EI}{E_{\mathrm{I}}}
\newcommand{\fI}{f_{\mathrm{I}}}
\newcommand{\RI}{R_{\mathrm{I}}}
\newcommand{\kI}{\kappa_{\mathrm{I}}}
\newcommand{\eIth}{e_{\mathrm{I}}^{\mathrm{th}}}

\newcommand{\Rs}{R_{\mathrm{s}}}
\newcommand{\fs}{f_{\mathrm{s}}}
\newcommand{\taus}{\tau_{\mathrm{s}}}
\newcommand{\taum}{\tau_{\mathrm{m}}}

\newcommand{\fin}{f_{\mathrm{in}}}
\newcommand{\iin}{i_{\mathrm{in}}}
\newcommand{\Esyn}{E_{\mathrm{syn}}}
\newcommand{\gsyn}{g_{\mathrm{syn}}}
\newcommand{\fC}{f_{\mathrm{C}}}

\newcommand{\gNa}{g_{\mathrm{Na}}}
\newcommand{\RNa}{R_{\mathrm{Na}}}
\newcommand{\ENa}{E_{\mathrm{Na}}}
\newcommand{\INa}{I_{\mathrm{Na}}}
\newcommand{\gK}{g_{\mathrm{K}}}
\newcommand{\RK}{R_{\mathrm{K}}}
\newcommand{\EK}{E_{\mathrm{K}}}
\newcommand{\IK}{I_{\mathrm{K}}}

\title{On Energy-Based Modelling for Fluidic Excitable Neurons}
\author{Robin~Wroblowski$^{1,*}$, Brent~Michiels$^2$, Benjamin~Gorissen$^2$, Rodolphe~Sepulchre$^{1,3}$
\thanks{The research leading to these results has received funding from the European Research Council under the Advanced ERC Grant Agreement SpikyControl n.101054323.}
\thanks{Email: {\itshape robin.wroblowski@kuleuven.be}, {\itshape brent.michiels@kuleuven.be}, {\itshape benjamin.gorissen@kuleuven.be}, {\itshape rodolphe.sepulchre@kuleuven.be}.}
\thanks{$^1$KU Leuven, Department of Electrical Engineering (ESAT), STADIUS Center for Dynamical Systems, Signal Processing and Data Analytics, KasteelPark Arenberg 10, 3001 Leuven, Belgium.}%
\thanks{$^2$KU Leuven, Department of Mechanical Engineering, 3001 Leuven, Belgium.}%
\thanks{$^3$Department of Engineering, University of Cambridge, Trumpington Street, Cambridge CB2 1PZ, United Kingdom.}%
}

\begin{document}
\maketitle

\begin{abstract}
    Motivated by the potential of pneumatic excitable devices in soft robotics, this paper develops an energy-based, domain-agnostic equivalent of the Hodgkin-Huxley modelling framework using bond graph theory. 
    The proposed model is used to guide the design of a pneumatic excitable neuron. 
    The potential of this approach is illustrated through the design of a pneumatic half-center oscillator, an essential building block for rhythmic neuromorphic behaviours.
\end{abstract}

\section{Introduction}

    Neuromorphic engineering draws inspiration from biological nervous systems to design efficient, event-driven hardware~\cite{meadNeuromorphicElectronicSystems1990}, achieving significant success in silicon microelectronics for abstract computation~\cite{christensen2022RoadmapNeuromorphic2022, sandamirskayaNeuromorphicComputingHardware2022}.
    
    

    A key potential of neuromorphic engineering is leveraging the multi-scale architecture of nervous systems across distinct physical domains, matching excitability timescales to their corresponding physical media~\cite{iidaTimescalesEmbodiedIntelligence2023}. 
    While electrical excitability operates on the millisecond scale~\cite{indiveriImportanceSpaceTime2019}, pneumatic excitability can be much slower, directly matching the temporal scale of mechanical motion.
    
    Currently, autonomous soft robotic control largely relies on continuous fluidic oscillators~\cite{rothemundSoftBistableValve2018, vanlaakeFluidicRelaxationOscillator2022}. 
    While effective for generating steady rhythmic patterns for locomotion, limit-cycle oscillators lack threshold-triggered excitability.
    In contrast, the design of excitable pneumatic devices has received little attention.
    This absence of a stable resting state prevents the implementation of event-based control, sensory-driven interrupts, and adaptive reflex loops~\cite{wroblowskiNeuromorphicDesignRhythmic2026a}.
    Integrate-and-fire models represent action potentials through threshold and reset rules, abstracting away the dynamics of spike generation~\cite{vanpottelberghBiophysicalIntegrateandFireModeling2021a}.
    Here, we instead pursue a continuous, conductance-based realisation of fluidic excitability.
    This choice avoids an unphysical reset and makes the roles of physical energy storage, supply, and state-dependent dissipation explicit.
    
    By grounding excitability in physical energy storage and dissipation rather than abstract state resets, conductance-based descriptions can be transformed into energy-based representations, such as bond graphs or port-Hamiltonian systems~\cite{duindamModelingControlComplex2009, vanderschaftPortHamiltonianModelingControl2020}. 
    This approach preserves energetic behaviour across mechanical, fluidic, and electrical domains alike~\cite{jeltsemaMemristivePortHamiltonianSystems2010}, providing a natural thermodynamic framework for controlling and simulating neuromorphic behaviour~\cite{rossControllingSimulatingSoft2016, chunEnergyBasedAbstractionSoft2023}.
    Previous efforts have successfully applied bond graph modelling to excitable membranes by explicitly capturing the underlying chemoelectrical energy transduction~\cite{gawthropBondGraphModelling2017a}.
    Additionally, bond graph and port-Hamiltonian methods have been directly integrated with memristor non-linear circuit elements to capture complex memory-dependent dynamics \cite{reading78141, jeltsemaMemristivePortHamiltonianSystems2010}.
    
    In this paper, we develop an energy-based model of an excitable neuron for the energy-aware development of excitable fluidic neurons.
    We first express the Hodgkin--Huxley circuit in terms of storage, supply, and state-dependent dissipative elements, and map this structure to chambers, pressure supplies, and variable restrictors.
    We then construct a reduced `EI' realisation following the mixed-feedback principle~\cite{burgerExcitabilityScalesUnifying2026,sepulchreControlScalesPositive2019}.
    Finally, pneumatic network simulations provide a numerical proof of concept through single-spike responses and alternating activity in a half center oscillator enabled by the post-inhibitory rebound property.

\section{Energetic Structure of the Hodgkin--Huxley Circuit}\label{sec:energetic_structure}

\begin{figure}[!t]
    \centering

 \begin{subfigure}[b]{\linewidth}
    \centering
    \resizebox{\linewidth}{!}{
    \begin{circuitikz}[
        >=latex,
        font=\large,
        european resistors, 
        sig/.style={line width=0.6pt, -{Latex[length=1.6mm]}, rounded corners=1.5mm}
    ]
        \definecolor{sigRed}{RGB}{213,11,11}

        \def\yT{3.5}   
        \def\yM{1.5}   
        \def\yB{0}     

        \def\xNa{1}    
        \def\xK{4}     
        \def\xL{7}     
        \def\xC{9.5}   
        \def\xIn{11.5} 

        \draw (0,\yT) -- (\xC,\yT);
        \node[above] at (0.3, \yT) {$V_{\mathrm{m}}$}; 
        
        \draw (0,\yB) -- (\xC,\yB);
        \draw (\xL,\yB) node[ground] {}; 

        \draw (\xNa,\yT) to[vR, l=$R_{\mathrm{Na}}$, *-] (\xNa,\yM)
                         to[battery1, l=$E_{\mathrm{Na}}$, -*] (\xNa,\yB);
        
        \draw (\xK,\yT) to[vR, l=$R_{\mathrm{K}}$, *-] (\xK,\yM)
                        to[battery1, l=$E_{\mathrm{K}}$, -*] (\xK,\yB);
                        
        \draw (\xL,\yT) to[R, l=$R_{\mathrm{L}}$, *-] (\xL,\yM)
                        to[battery1, l=$E_{\mathrm{L}}$, -*] (\xL,\yB);

        \draw (\xC,\yT) to[C, l=$C_{\mathrm{m}}$, *-*] (\xC,\yB);

        \draw (\xIn,\yT) to[american current source, l_=$i_{\mathrm{in}}$] (\xC,\yT);

        \coordinate (TapNa) at (\xNa-0.9, \yT);
        \draw[sig, sigRed] (TapNa) -- (\xNa-0.9, 2.5) node[midway, right] {$e_{\mathrm{m}}$} -- (\xNa-0.3, 2.5);
        
        \coordinate (TapK) at (\xK-0.9, \yT);
        \draw[sig, sigRed] (TapK) -- (\xK-0.9, 2.5) node[midway, right] {$e_{\mathrm{m}}$} -- (\xK-0.3, 2.5);

    \end{circuitikz}
    }
    \caption{Electrical circuit representation of the Hodgkin-Huxley model.}
    \label{fig:hh-circuit}
\end{subfigure}
    
    \vspace{0.8em} 

    \begin{subfigure}[b]{\linewidth}
        \centering
        \resizebox{\linewidth}{!}{
        \begin{tikzpicture}[
            >=latex,
            el/.style={draw, rectangle, line width=0.5pt, minimum width=1.0cm, minimum height=0.55cm, inner sep=2pt, font=\small},
            jn/.style={inner sep=2pt, minimum size=0.5cm, font=\small\bfseries},
            lbl/.style={font=\scriptsize, inner sep=1.5pt},
            bondCH/.style={line width=0.6pt, -{Straight Barb[length=2mm, width=1.6mm, left] Bar[width=3mm, line width=0.6pt]}},
            bondCT/.style={line width=0.6pt, {Bar[width=3mm, line width=0.6pt]}-{Straight Barb[length=2mm, width=1.6mm, left]}},
            sig/.style={line width=0.6pt, -{Latex[length=1.6mm]}, rounded corners=1.5mm}
        ]
            \definecolor{sigRed}{RGB}{213,11,11}
        
            \def\xL{2.4}   \def\xM{-0.6}  \def\xF{5.4}
            \def\yJtop{1.2} \def\yRtop{1.2}

            \def\xNa{1.1}   \def\xK{3.7}  
            \def\yJ{-1.2}   \def\yR{-2.5}
        
            \node[draw, line width=0.5pt, fill=black!6, minimum width=7.2cm, minimum height=0.38cm,
                  inner sep=0pt, font=\small\bfseries] (J0) at (2.4, 0) {0};
            \node[font=\small, left=2pt] at (J0.west) {$e_{\mathrm{m}}$};
        
            \node[jn] (J1L) at (\xL, \yJtop) {1};
            \node[el] (RL)  at (\xL-1.65, \yRtop) {$\mathrm{R}\!:\!R_{\mathrm{L}}$};
            \node[el] (SeL) at (\xL+1.65, \yJtop) {$\mathrm{Se}\!:\!E_{\mathrm{L}}$};
        
            \node[el] (Cm)  at (\xM, \yJtop) {$\mathrm{C}\!:\!C_{\mathrm{m}}$};
            \node[el] (Sf)  at (\xF, \yJtop) {$\mathrm{Sf}\!:\!f_{\mathrm{in}}$};
        
            \draw[bondCH, shorten >=2pt] (J0.north -| J1L) -- (J1L)
                node[lbl, left, pos=0.4] {$e_{\mathrm{m}}$} node[lbl, right, pos=0.4] {$f_{\mathrm{L}}$};
            \draw[bondCH, shorten >=2pt] (J1L) -- (RL)
                node[lbl, above, pos=0.4] {$e_{R_{\mathrm{L}}}$} node[lbl, below, pos=0.4] {$f_{\mathrm{L}}$};
            \draw[bondCT] (J1L) -- (SeL)
                node[lbl, above, pos=0.35] {$E_{\mathrm{L}}$} node[lbl, below=0pt, pos=0.35] {$f_{\mathrm{L}}$};
        
            \draw[bondCT, shorten <=2pt] (J0.north -| Cm) -- (Cm)
                node[lbl, left, pos=0.4] {$e_{\mathrm{m}}$} node[lbl, right, pos=0.4] {$f_{\mathrm{m}}$};
        
            \draw[bondCT, shorten <=2pt] (Sf) -- (J0.north -| Sf)
                node[lbl, left, pos=0.4] {$e_{\mathrm{m}}$} node[lbl, right, pos=0.4] {$f_{\mathrm{in}}$};
        
            \node[jn] (J1Na) at (\xNa, \yJ) {1};
            \node[el, minimum width=1.8cm] (MRNa) at (\xNa, \yR) {$\mathrm{MR}\!:\!R_{\mathrm{Na}}$};
            \node[el] (SeNa) at (\xNa-1.8, \yJ) {$\mathrm{Se}\!:\!E_{\mathrm{Na}}$};
        
            \node[jn] (J1K) at (\xK, \yJ) {1};
            \node[el, minimum width=1.8cm] (MRK)  at (\xK, \yR) {$\mathrm{MR}\!:\!R_{\mathrm{K}}$};
            \node[el] (SeK)  at (\xK+1.8, \yJ) {$\mathrm{Se}\!:\!E_{\mathrm{K}}$};
        
            \foreach \j/\f in {J1Na/f_{\mathrm{Na}}, J1K/f_{\mathrm{K}}} {
                \draw[bondCH] (J0.south -| \j) -- (\j)
                    node[lbl, left, pos=0.4] {$e_{\mathrm{m}}$} node[lbl, right, pos=0.4] {$\f$};
            }
            
            \draw[bondCH, shorten >=2pt] (J1Na) -- (MRNa) node[lbl, left, pos=0.3] {$e_{R_{\mathrm{Na}}}$} node[lbl, right, pos=0.3] {$f_{\mathrm{Na}}$};
            \draw[bondCH, shorten >=2pt] (J1K) -- (MRK)   node[lbl, left, pos=0.3] {$e_{R_{\mathrm{K}}}$} node[lbl, right, pos=0.3] {$f_{\mathrm{K}}$};
            
            \draw[bondCT] (J1Na) -- (SeNa) node[lbl, above, pos=0.45] {$E_{\mathrm{Na}}$} node[lbl, below, pos=0.45] {$f_{\mathrm{Na}}$};
            \draw[bondCT] (J1K) -- (SeK)   node[lbl, above, pos=0.45] {$E_{\mathrm{K}}$} node[lbl, below, pos=0.45] {$f_{\mathrm{K}}$};
        
            \fill[sigRed] (\xNa+0.65, -0.19) circle (1.2pt);
            \draw[sig, sigRed] (\xNa+0.65, -0.19) -- (\xNa+0.65, -2.15)
                node[lbl, right, midway, text=sigRed] {$V_{\mathrm{m}}$}; 

            \fill[sigRed] (\xK-0.65, -0.19) circle (1.2pt);
            \draw[sig, sigRed] (\xK-0.65, -0.19) -- (\xK-0.65, -2.15)
                node[lbl, left, midway, text=sigRed] {$V_{\mathrm{m}}$}; 
        
        \end{tikzpicture}
        }
        \caption{Domain-agnostic bond graph representation.}
        \label{fig:hh-bondgraph}
    \end{subfigure}

    \caption{Equivalent representations of the Hodgkin--Huxley model of neuronal excitability: (a) canonical electrical circuit and (b) domain-independent bond graph model based on energy storage ($C_m$), supply ($E_x$), and dissipation ($R_x$). Red arrows indicate the modulation signals ($e_{\mathrm{m}}$/$V_{\mathrm{m}}$), i.e. the state-dependence of the modulated resistances $R_{\mathrm{Na}}(V_m)$ and $R_{\mathrm{K}}(V_m)$.}
    \label{fig:hh-models}
\end{figure}
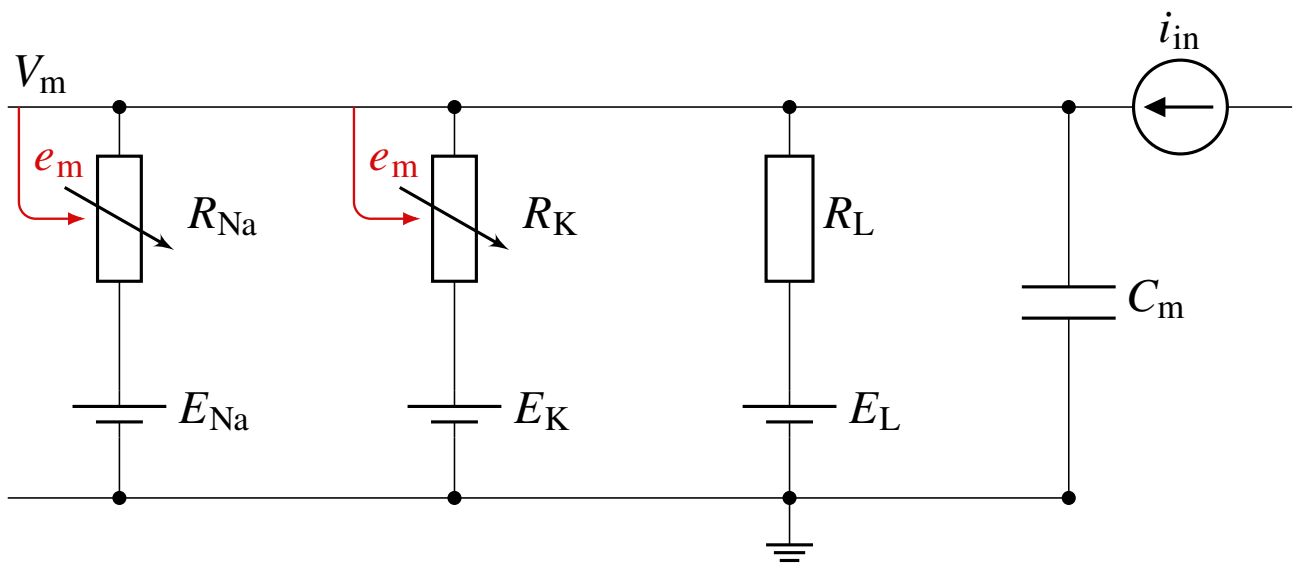
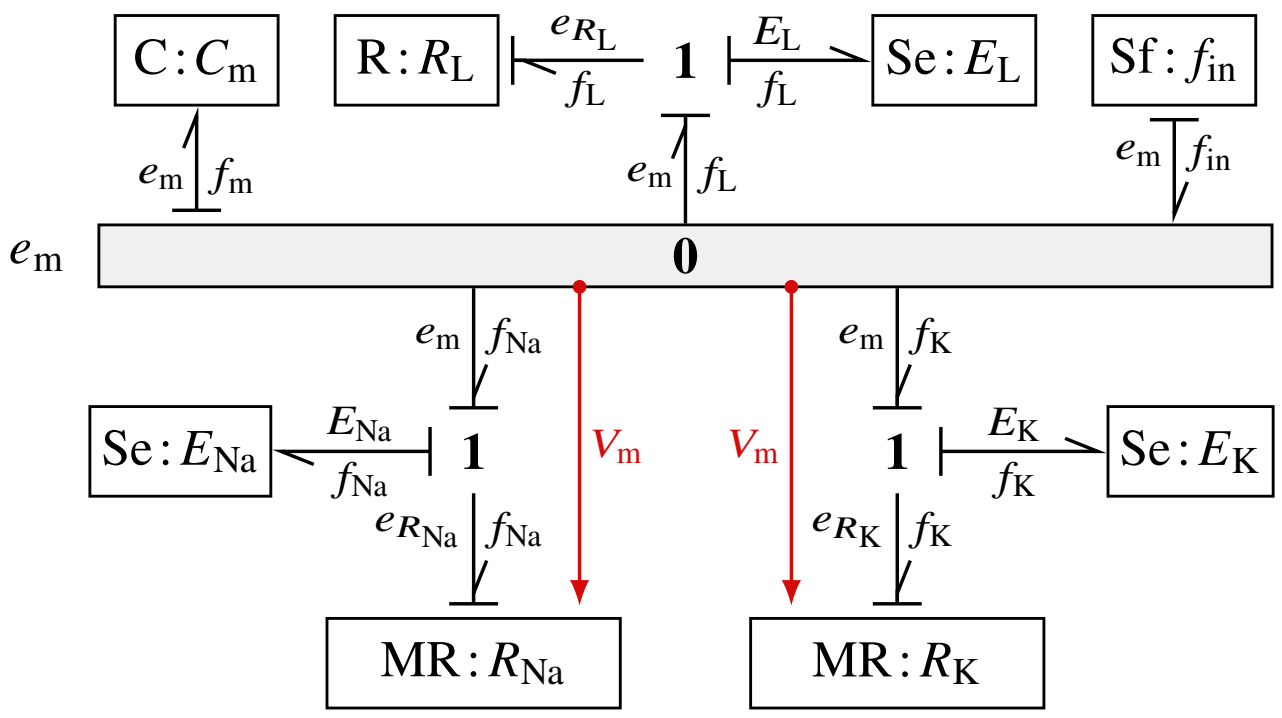

    The seminal work of Hodgkin and Huxley~\cite{hodgkinQuantitativeDescriptionMembrane1952} laid the foundation for conductance-based neuronal modelling by describing the ionic mechanisms underlying action potential generation as an equivalent electrical circuit.
    While this circuit translates directly into a state-space model for mathematical simulation, viewing the system purely through its differential equations obscures the underlying energetic topology. 
    Retaining the physical Hodgkin--Huxley circuit (Fig.~\ref{fig:hh-circuit}) as the primary model makes the energy storage, supply, and state-dependent dissipation explicit.
    
    Topologically, the circuit consists of the parallel interconnection of five branches across a shared membrane voltage $\Vm$: a capacitive branch ($\Cm$), an external current branch ($\iin$), a passive leak branch ($\RL, \EL$), and two active ionic branches representing the sodium current $\INa$ ($\RNa, \ENa$) and potassium current $\IK$ ($\RK, \EK$).
    Inspecting the constitutive components of these branches reveals that the entire circuit is constructed from only three fundamental classes of physical elements, each defined by a distinct energetic role in terms of generalised effort $e(t)$ (voltage) and flow $f(t)$ (current), whose product defines instantaneous power $$\mathcal{P}(t) = e(t)f(t).$$
    
    \noindent\textbf{Remarks on notation:} 
    Bond graphs provide a domain-agnostic graphical formalism for modelling the energy dynamics of physical systems~\cite{karnoppSystemDynamicsModeling2012}. 
    Graphically, power exchange between components is depicted by bonds (half-arrows) pointing in the direction of positive power flow, carrying power-conjugate effort ($e$) and flow ($f$) variables. 
    Network topology is governed by two multi-port nodes: $0$-junctions (parallel nodes with common effort) and $1$-junctions (series nodes with common flow).
    Following standard conventions, physical components are denoted by their functional type (e.g., `$\text{C}$' for storage, `$\text{R}$' for dissipation) and a specific identifier (e.g., `$\text{C}{:}\Cm$`, `$\text{R}{:}\RL$`). 
    Additionally, while neuroscience typically refers to ionic channels by their conductances ($\gL, \gNa, \gK$), we represent them by their equivalent reciprocal resistances ($R_x = 1/g_x$) to align with standard energetic dissipators.
    \par\vspace{0.5em}
    
    \subsection{Capacitors as energy-storage elements} 
        The membrane capacitor $\Cm$ acts as the circuit's primary and only energy storage element.
        Integrating the capacitive flow $\fC(t)$ accumulates generalised displacement (charge $\qm(t) = \int \fC(t)\,\mathrm{d}t$), thereby storing electrostatic energy via the Hamiltonian $$\mathcal{H}(\qm) = \frac{1}{2\Cm}\qm^2.$$
        The shared membrane effort is then recovered as the co-energy variable $$\emem(t) = \frac{\partial \mathcal{H}}{\partial \qm} = \frac{\qm(t)}{\Cm}.$$

    \subsection{Batteries and current inputs as energy-supply elements} 
        The generation of action potentials relies on an active energy supply, realised through ideal sources that represent trans-membrane electrochemical gradients.
        Internally, this energy supply is delivered by reversal potentials ($\ENa, \EK, \EL$) acting as constant effort sources ($\text{Se}$) in series with each branch's memductance.  
        Conversely, the injected current $\iin(t)$ acts as an external energy supply represented by an ideal flow source ($\text{Sf}$).
        Both types of source prescribe their specified input power variable independently of the conjugate power variable.
        
    \subsection{Memductors as state-dependent dissipative elements} 
        Ionic channels regulate charge flow through the linear leak conductance $\gL$ (element `$\mathrm{R}{:}\RL$') and the nonlinear, voltage-gated conductances $\gNa$ and $\gK$. 
        Energetically, $\gNa$ and $\gK$ are not active power generators, but \textit{memductors}, i.e., dissipative elements with internal memory~\cite{chuaHodgkinHuxleyAxon2012}. 
        
        While previous thermodynamic bond graphs explicitly model chemoelectrical Markov gating states~\cite{gawthropBondGraphModelling2017a}, this microscopic detail is unnecessary for our purpose. Translating these macroscopic dynamics into the fluidic domain only requires capturing the state-dependent dissipation, not the underlying biological kinetics.        
        Mathematically, these mem-elements can be equivalently portrayed as modulated resistors ($\mathrm{MR}$) whose state-dependent resistances $R(\cdot)$ vary over time. 
        While memristive elements exist in bond graph theory, standard formulations restrict them to linear displacement-modulation \cite{osterMemristorNewBond1972}. 
        Using explicitly parametrised $\mathrm{MR}$ elements avoids the complex extensions required for non-linear memristors \cite{reading78141} while still capturing the gating nonlinearities essential for our fluidic translation.
        
        Because $g_x(\cdot)\ge0$, each modulated resistor is passive at its power port, with instantaneous dissipation \[\mathcal{P}_x(t)=e_x(t)f_x(t)=g_x(\cdot)e_x(t)^2\ge0.\]
        The gating variables modulate the conductance through ideal information signals; the energetic cost associated with their physical implementation is not represented at these control ports.
   
    \subsection{Bond graph representation}    
        Because storage ($\text{C}$), supply ($\text{Se}$, $\text{Sf}$), and (state-dependent) dissipation ($\text{R}, \text{MR}$) are defined purely via generalised effort and flow, this energy-based description is domain-agnostic.  
        The Hodgkin--Huxley circuit can therefore be generalised directly into the bond graph shown in Fig.~\ref{fig:hh-bondgraph}.
        Here, a central $0$-junction enforces flow continuity ($\sum f = 0$, Kirchhoff's Current Law) at the common membrane effort $\emem$, while $1$-junctions enforce common branch flows $f_x$ and effort drop balance ($\Delta e_x = \emem - E_x$, Kirchhoff's Voltage Law) across each series ionic branch $x \in \{\text{Na}, \text{K}, \mathrm{L}\}$. 
        This bond graph can be translated directly into the fluidic domain.
    
\section{Translation into the fluidic domain}\label{sec:fluidic_domain}

    The bond graph representation in Section~\ref{sec:energetic_structure} provides a domain-agnostic description of excitable dynamics governed by the power-conjugate variables of effort and flow.
    To physically realise our goal of continuous fluidic excitability for soft robotic control, we map this structure into the fluidic domain by adopting the standard fluidic impedance analogy:
    \begin{equation}
        e(t) = P(t), \qquad f(t) = Q(t),
        \label{eq:fluid_mapping}
    \end{equation}
    where $P$ is gauge pressure relative to atmospheric reference and $Q$ is volumetric flow rate.
    
    Unlike mechanical analogies, which require choosing between preservation of topology or generalised impedance~\cite{jeltsemaMultidomainModelingNonlinear2009a}, the fluidic realisation of the bond graph maintains the standard across/through variable assignment~\cite{karnoppSystemDynamicsModeling2012}, thereby conserving both the underlying network topology and generalised impedance.
    Generalised effort corresponds directly to gauge pressure and acts as the across-variable, satisfying path independence around closed loops, $\sum \Delta P = 0$ ($\Delta P$ being the pressure drop across an element).
    Consequently, series fluidic interconnections map directly onto $1$-junctions, where the effort drops are additive. Likewise, generalised flow corresponds to volumetric flow rate and acts as the through-variable, satisfying continuity at network nodes, $\sum Q = 0$. 
    Fluidic branches that share a common pressure drop divide the flow between them, mapping directly onto $0$-junctions.
    
    This energetic mapping allows the bond graph components to be translated one-to-one into physical fluidic hardware:
    \begin{itemize}
        \item \textit{Generalised storage ($\text{C}$)}: The capacitive energy storage of the membrane ($\Cm$) is realised by elastic accumulators or rigid fluidic chambers, whose compliance is defined by $C = \mathrm{d}V/\mathrm{d}P$.
        \item \textit{Modulated dissipation ($\text{MR}$)}: The state-dependent ionic conductances ($\gNa, \gK$) are realised by pressure-controlled valves or variable fluidic restrictors. 
        Their hydraulic resistance is modulated by a control pressure.
        We represent this modulation through an ideal, unloaded control signal, neglecting control-port loading.
        The restriction remains dissipative at its flow ports.
        \item \textit{Constant effort sources ($\text{Se}$)}: The reversal potentials ($\ENa, \EK, \EL$) map directly to constant-pressure supply manifolds and atmospheric exhaust ports. 
        This avoids the need for dedicated flow sources or active metering pumps (which would be required in the dual mapping of the mobility analogy).
        \item \textit{Embodied actuation coupling}: Because the generalised membrane effort maps directly to physical fluid pressure ($\emem = P_m$), the internal neuronal state couples statically to mechanical output. 
        Depending on the actuator morphology, this internal pressure transduces into either linear forces ($F \propto P_m$) or bending moments ($\tau \propto P_m$).
    \end{itemize}

    \subsection*{Lumped pneumatic regime}\label{sec:fluidic_domain:lumped}
        
        We adopt a lumped pneumatic approximation with spatially uniform chamber pressures, negligible fluid inertance, and quasi-steady resistive flow laws~\cite{karnoppSystemDynamicsModeling2012}.
        Gas compressibility and quasi-static chamber elasticity are represented by an effective capacitance for each chamber, while flow nonlinearities enter through the branch constitutive laws.
        This yields a nonlinear RC network, neglecting additional inertial and thermal dynamics over the timescales of interest.
        
\section{Implementation of gated memductive element}
    
    The intention of the Hodgkin-Huxley model was to accurately capture the action potential dynamics of the squid giant axon.
    Because of this, it contains features that complicate physical realisation, such as polynomial gating exponents and voltage-dependent time constants within the ionic currents.
    To reduce the complexity of the fluidic implementation, we instead use a reduced system, which retains the essential mechanisms underlying excitability while discarding details that fine-tune the spike waveform.

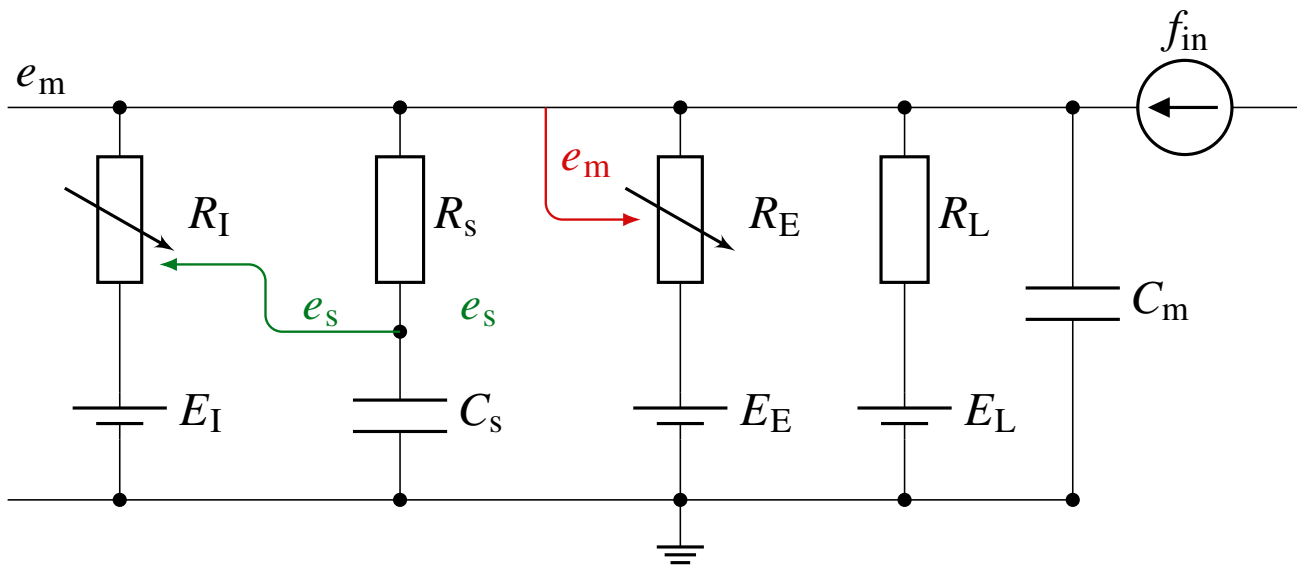
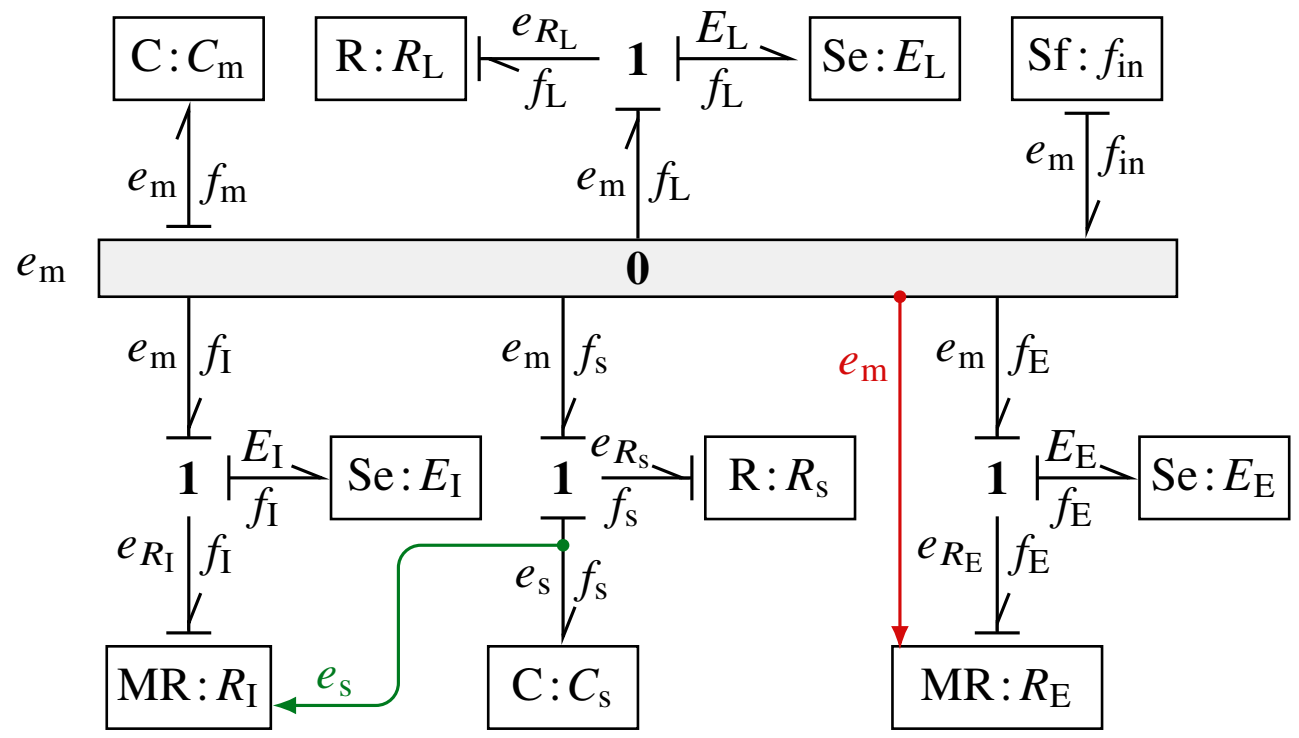
\begin{figure}[!t]
    \centering

    \begin{subfigure}[b]{\linewidth}
        \centering
        \resizebox{\linewidth}{!}{
        \begin{circuitikz}[
            font=\large,
            >=latex,
            european resistors, 
            sig/.style={line width=0.6pt, -{Latex[length=1.6mm]}, rounded corners=1.5mm}
        ]
            \definecolor{sigGreen}{RGB}{0,122,31}
            \definecolor{sigRed}{RGB}{213,11,11}
        
            \def\yT{3.5}   
            \def\yM{1.5}   
            \def\yB{0}     
        
            \def\xI{1}     
            \def\xS{3.5}   
            \def\xE{6}     
            \def\xL{8}     
            \def\xC{9.5}   
            \def\xIn{11.5} 
        
            \draw (0,\yT) -- (\xC,\yT);
            \node[above] at (0.3, \yT) {$e_{\mathrm{m}}$};
            
            \draw (0,\yB) -- (\xC,\yB);
            \draw (\xE,\yB) node[ground] {}; 
        
            \draw (\xI,\yT) to[vR, l=$R_{\mathrm{I}}$, *-] (\xI,\yM)
                            to[battery1, l=$E_{\mathrm{I}}$, -*] (\xI,\yB);
            
            \draw (\xS,\yT) to[R, l=$R_{\mathrm{s}}$, *-*] (\xS,\yM) coordinate (esNode)
                            to[C, l=$C_{\mathrm{s}}$, -*] (\xS,\yB);
                            
            \draw (\xE,\yT) to[vR, l=$R_{\mathrm{E}}$, *-] (\xE,\yM)
                            to[battery1, l=$E_{\mathrm{E}}$, -*] (\xE,\yB);
                            
            \draw (\xL,\yT) to[R, l=$R_{\mathrm{L}}$, *-] (\xL,\yM)
                            to[battery1, l=$E_{\mathrm{L}}$, -*] (\xL,\yB);
        
            \draw (\xC,\yT) to[C, l=$C_{\mathrm{m}}$, *-*] (\xC,\yB);
        
            \draw (\xIn,\yT) to[american current source, l_=$f_{\mathrm{in}}$] (\xC,\yT);
            
            \node[xshift=-20pt, yshift=5pt, text=sigGreen] at (esNode) {$e_{\mathrm{s}}$};
            \node[xshift=20pt, yshift=5pt, text=sigGreen] at (esNode) {$e_{\mathrm{s}}$};
        
            \draw[sig, sigGreen] (esNode) -- ++(-1.2,0.) |- (\xI+0.35, 2.1);
            
        
            \coordinate (emTap) at (\xE-1.2, \yT);
            \draw[sig, sigRed] (emTap) -- (\xE-1.2, 2.5) node[midway, right] {$e_{\mathrm{m}}$} -- (\xE-0.35, 2.5);
        
        \end{circuitikz}
        }
        \caption{Lumped electrical circuit with modulated resistors representing the memristive elements.}
        \label{fig:EI-circuit}
    \end{subfigure}

    \vspace{0.8em} 

    \begin{subfigure}[b]{\linewidth}
        \centering
        \resizebox{\linewidth}{!}{
        \begin{tikzpicture}[
            >=latex,
            el/.style={draw, rectangle, line width=0.5pt, minimum width=1.0cm, minimum height=0.55cm, inner sep=2pt, font=\small},
            jn/.style={inner sep=2pt, minimum size=0.5cm, font=\small\bfseries},
            lbl/.style={font=\small, inner sep=1.5pt},
            bondCH/.style={line width=0.6pt, -{Straight Barb[length=2mm, width=1.6mm, left] Bar[width=3mm, line width=0.6pt]}},
            bondCT/.style={line width=0.6pt, {Bar[width=3mm, line width=0.6pt]}-{Straight Barb[length=2mm, width=1.6mm, left]}},
            sig/.style={line width=0.6pt, -{Latex[length=1.6mm]}, rounded corners=1.5mm, font=\small}
        ]
            \definecolor{sigGreen}{RGB}{0,122,31}
            \definecolor{sigRed}{RGB}{213,11,11}
        
            \def\xL{2.4}   \def\xM{-0.6}  \def\xF{5.4}
            \def\yJtop{1.4} \def\yRtop{1.4}

            \def\xI{-0.6}   \def\xS{1.9}  \def\xE{4.8}
            \def\yJ{-1.4}   \def\yR{-2.8}
        
            \node[draw, line width=0.5pt, fill=black!6, minimum width=7.2cm, minimum height=0.38cm,
                  inner sep=0pt, font=\small\bfseries] (J0) at (2.4, 0) {0};
            \node[font=\small, left=2pt] at (J0.west) {$e_{\mathrm{m}}$};
        
            \node[jn] (J1L) at (\xL, \yJtop) {1};
            \node[el] (RL)  at (\xL-1.65, \yRtop) {$\mathrm{R}\!:\!R_{\mathrm{L}}$};
            \node[el] (SeL) at (\xL+1.65, \yJtop) {$\mathrm{Se}\!:\!E_{\mathrm{L}}$};
        
            \node[el] (Cm)  at (\xM, \yJtop) {$\mathrm{C}\!:\!C_{\mathrm{m}}$};
            \node[el] (Sf)  at (\xF, \yJtop) {$\mathrm{Sf}\!:\!f_{\mathrm{in}}$};
        
            \draw[bondCH, shorten >=2pt] (J0.north -| J1L) -- (J1L)
                node[lbl, left, pos=0.4] {$e_{\mathrm{m}}$} node[lbl, right, pos=0.4] {$f_{\mathrm{L}}$};
            \draw[bondCH, shorten >=2pt] (J1L) -- (RL)
                node[lbl, above, pos=0.4] {$e_{R_{\mathrm{L}}}$} node[lbl, below, pos=0.4] {$f_{\mathrm{L}}$};
            \draw[bondCT, shorten >=2pt] (J1L) -- (SeL)
                node[lbl, above, pos=0.35] {$E_{\mathrm{L}}$} node[lbl, below=0pt, pos=0.35] {$f_{\mathrm{L}}$};
        
            \draw[bondCT, shorten <=2pt] (J0.north -| Cm) -- (Cm)
                node[lbl, left, pos=0.4] {$e_{\mathrm{m}}$} node[lbl, right, pos=0.4] {$f_{\mathrm{m}}$};
        
            \draw[bondCT, shorten <=2pt] (Sf) -- (J0.north -| Sf)
                node[lbl, left, pos=0.4] {$e_{\mathrm{m}}$} node[lbl, right, pos=0.4] {$f_{\mathrm{in}}$};
        
            \node[jn] (J1I) at (\xI, \yJ) {1};
            \node[el] (MRI) at (\xI, \yR) {$\mathrm{MR}\!:\!R_{\mathrm{I}}$};
            \node[el] (SeI) at (\xI+1.45, \yJ) {$\mathrm{Se}\!:\!E_{\mathrm{I}}$};
        
            \node[jn] (J1S) at (\xS, \yJ) {1};
            \node[el] (Cs)  at (\xS, \yR) {$\mathrm{C}\!:\!C_{\mathrm{s}}$};
            \node[el] (Rs)  at (\xS+1.45, \yJ) {$\mathrm{R}\!:\!R_{\mathrm{s}}$};
        
            \node[jn] (J1E) at (\xE, \yJ) {1};
            \node[el, minimum width=1.4cm] (MRE) at (\xE, \yR) {$\mathrm{MR}\!:\!R_{\mathrm{E}}$};
            \node[el] (SeE) at (\xE+1.45, \yJ) {$\mathrm{Se}\!:\!E_{\mathrm{E}}$};
        
            \foreach \j/\f in {J1I/f_{\mathrm{I}}, J1S/f_{\mathrm{s}}, J1E/f_{\mathrm{E}}} {
                \draw[bondCH] (J0.south -| \j) -- (\j)
                    node[lbl, left, pos=0.4] {$e_{\mathrm{m}}$} node[lbl, right, pos=0.4] {$\f$};
            }
            
            \draw[bondCH, shorten >=2pt] (J1I) -- (MRI) node[lbl, left, pos=0.3] {$e_{R_{\mathrm{I}}}$} node[lbl, right, pos=0.3] {$f_{\mathrm{I}}$};
            \draw[bondCH, shorten >=2pt] (J1E) -- (MRE) node[lbl, left, pos=0.3] {$e_{R_{\mathrm{E}}}$} node[lbl, right, pos=0.3] {$f_{\mathrm{E}}$};
            
            \draw[bondCT] (J1I) -- (SeI) node[lbl, above, pos=0.35] {$E_{\mathrm{I}}$} node[lbl, below, pos=0.35] {$f_{\mathrm{I}}$};
            \draw[bondCT] (J1E) -- (SeE) node[lbl, above, pos=0.35] {$E_{\mathrm{E}}$} node[lbl, below, pos=0.35] {$f_{\mathrm{E}}$};
            
            \draw[bondCH, shorten >=2pt] (J1S) -- (Rs)  node[lbl, above, pos=0.2] {$e_{R_{\mathrm{s}}}$} node[lbl, below, pos=0.2] {$f_{\mathrm{s}}$};
            \draw[bondCT] (J1S) -- (Cs)  node[lbl, left, pos=0.5]  {$e_{\mathrm{s}}$} node[lbl, right, pos=0.5] {$f_{\mathrm{s}}$};
        
            \coordinate (esTap) at (1.9, -1.85); 
            \fill[sigGreen] (esTap) circle (1.2pt);
            \draw[sig, sigGreen] (esTap) -- (0.8, -1.85) |- ([yshift=-1.2mm]MRI.east)
                node[lbl, pos=0.75, above, text=sigGreen] {$e_{\mathrm{s}}$};
        
            \fill[sigRed] (4.15, -0.19) circle (1.2pt);
            \draw[sig, sigRed] (4.15, -0.19) -- (4.15, -2.55)
                node[lbl, left, pos=0.2, text=sigRed] {$e_{\mathrm{m}}$}; 
        
        \end{tikzpicture}
        }
        \caption{Bond graph with modulated resistors representing the memristive elements.}
        \label{fig:EI-bond}
    \end{subfigure}

    \caption{Circuit (a) and bond graph (b) representations of the EI neuron model. Modulation signals are indicated by red ($e_{\mathrm{m}}$) and green ($e_{\mathrm{s}}$) arrows.}
    \label{fig:EI-combined}
\end{figure}

    The reduced model, referred to here as the \textit{EI} (Excitatory-Inhibitory) neuron model~\cite{burgerExcitabilityScalesUnifying2026} and depicted as an equivalent circuit in Fig.~\ref{fig:EI-circuit}, consists of two dynamical states, representing the membrane effort $\emem$ and a slow recovery variable $\es$. 
    The EI model discards both the voltage dependence of the time constants and the power exponents of the gating variables.
    Furthermore, fast excitation is represented by a static sigmoidal function of $\emem$, while delayed inhibition is governed by a static sigmoidal function of $\es$, a first-order low-pass filtered version of $\emem$.    
    This structure follows the mixed-feedback motif~\cite{sepulchreControlScalesPositive2019}, which separates fast positive feedback (excitation) from slower negative feedback (inhibition).

    Rather than using electrical variables ($V$ and $I$), we formulate the EI equations directly in terms of generalised effort ($e$) and flow ($f$) variables:
    \begin{equation}
    \left\{
    \begin{aligned}
        \Cm \dot{e}_{\mathrm{m}} &= \fin(t) - \fL - \fE - \fI - \fs, \\
        \Cs \dot{e}_{\mathrm{s}} &= \fs,
    \end{aligned}
    \right.
    \label{eq:ei_dynamics}
    \end{equation}
    \begin{align}
        \fs  &= \frac{\emem - \es}{\Rs}, \label{eq:ei_fs} \\
        f_x  &= g_x \: (\emem - E_x) \quad x \in \{\mathrm{E}, \mathrm{I}, \mathrm{L}\}, \label{eq:ei_fx} \\
        \gE  &= \gEbar\:  \sigma\bigl(\kE (\emem - \eEth)\bigr), \label{eq:ei_ge} \\
        \gI  &= \gIbar \: \sigma\bigl(\kI (\es - \eIth)\bigr), \label{eq:ei_gi} \\
        \gL  &= \gLbar \label{eq:ei_gl}
    \end{align}
    where the activation function $\sigma(\cdot)$ is the logistic sigmoid:
    \begin{equation}
        \sigma(z) = \frac{1}{1 + \exp(-z)}.
        \label{eq:sigmoid}
    \end{equation}

    Here, $\emem$ is the effort of the membrane storage $\Cm$, and $\es$ is the effort across the slow recovery storage $\Cs$.
    The input flow is denoted by $\fin(t)$.
    The flows $\fE$, $\fI$, and $\fL$ represent the excitatory, inhibitory, and leak branches, driven by constant reversal effort sources ordered as $\EI \le \EL < \EE$. 

    The constants $\gEbar$, $\gIbar$, and $\gLbar$ represent the maximum conductances for each memductance branch. 
    The excitatory conductance $\gE$ is controlled by the membrane effort $\emem$, with sensitivity $\kE$ and threshold $\eEth$.
    The inhibitory conductance $\gI$ is controlled by the recovery effort $\es$, with sensitivity $\kI$ and threshold $\eIth$.
    The series resistance $\Rs$ sets the recovery time constant $\taus=\Rs\Cs$.
    
    The recovery variable $\es$ is implemented as a passive $\text{RC}$ low-pass filter rather than as an abstract integrated state. 
    In the reduced bond graph of Fig.~\ref{fig:EI-bond}, the storage element $\text{C}{:}\Cs$ receives the flow $\fs$ from the membrane node through the linear dissipator $\text{R}{:}\Rs$. 
    The ionic branches are consolidated into two memductors: a fast excitatory element, $\text{MR}{:}\RE$, modulated by $\emem$, and a slow inhibitory element, $\text{MR}{:}\RI$, modulated by $\es$. 
    Both storage elements have integral causality, and the interconnection introduces no algebraic loops of causality.
    The resulting equations therefore define an explicit two-state model with state-dependent resistive branches.

    \subsection*{Energy balance}

        For the constant-capacitance EI model, define the storage function
        \begin{equation}
            \mathcal{H}
            =\frac{1}{2}\Cm\emem^2
            +\frac{1}{2}\Cs\es^2.
            \label{eq:ei_storage}
        \end{equation}
        Using~\eqref{eq:ei_dynamics}--\eqref{eq:ei_fx}, its derivative is
        \begin{equation}
        \begin{aligned}
            \dot{\mathcal{H}}
            ={}&\emem\fin
            -\sum_{x\in\{\mathrm{E},\mathrm{I},\mathrm{L}\}} E_x f_x\\
            &-\sum_{x\in\{\mathrm{E},\mathrm{I},\mathrm{L}\}}
                g_x(\emem-E_x)^2
            -\frac{(\emem-\es)^2}{\Rs}.
        \end{aligned}
        \label{eq:ei_energy_balance}
        \end{equation}
        The first two terms describe exchange with the input and reversal sources, while the remaining terms are non-negative dissipation with a minus sign.
        Excitability thus arises from state-dependent access to the effort supplies, without requiring active power generation by the resistive elements.
    
\section{Simulation results}
    We simulate the excitable fluidic network using the Python package `FONSim'~\cite{baeyensFONSPythonFramework2023a} in a compressible, lumped pneumatic regime operating around atmospheric reference ($P_{\mathrm{atm}} = 101.3\,\text{kPa}$).

    \subsection{Network parametrisation and simulation setup}
        We map the network onto standard FONSim components: rigid container elements model the storage chambers ($\Cm, \Cs$), uniform tube elements define the linear dissipative pathways ($\RL, \Rs$), and ideal pressure ports supply the constant reversal levels ($E_x$). 
        To avoid vacuum generation in soft robotics, the pressure scale is shifted so that the inhibitory exhaust rests at atmospheric pressure ($\EI = 0\,\text{bar}$ gauge), the excitatory supply is set to a regulated line pressure of $\EE = 1.0\,\text{bar}$ gauge, and the leak potential is biased to $\EL = 0.2\,\text{bar}$ gauge. 
        Input stimuli are delivered as transient mass-flow pulses $\fin(t)$, reflecting event-driven sensory inputs rather than continuous exogenous supplies.

        The storage chambers are assigned internal volumes of $V_{\mathrm{m}} = V_{\mathrm{s}} = 1\,\text{mL}$, deriving their capacitance purely from gas compressibility. 
        Fluidic dissipation is realised using restrictor tubes of reference length $L = 1.0\,\text{m}$ with effective inner diameters of $d_{\mathrm{L}} = 1.0\,\text{mm}$ for the passive leak and $d_{\mathrm{s}} = 0.3\,\text{mm}$ for the slow recovery pathway. 
        Together, these yield a passive membrane time constant of $\taum \approx 7\,\text{ms}$ and a slow recovery time constant of $\taus \approx 0.9\,\text{s}$. 
        This timescale separation ($\taum / \taus \sim 10^{-2}$) yields sharp spikes alongside slow recovery suitable for locomotion.
        
        The active excitatory and inhibitory branches use larger effective diameters ($d_{\mathrm{E}} = 2.0\,\text{mm}$, $d_{\mathrm{I}} = 3.0\,\text{mm}$) with sigmoidal thresholds centred at $p_{\mathrm{th,E}} = p_{\mathrm{th,I}} = 0.45\,\text{bar}$ gauge and sensitivities $a_{\mathrm{E}} = 0.3\,\text{kPa}^{-1}$, $a_{\mathrm{I}} = 0.2\,\text{kPa}^{-1}$. 
        For the half-center oscillator, mutual inhibition is mediated by a synaptic restrictor ($d_{\mathrm{syn}} = 6.0\,\text{mm}$, $p_{\mathrm{th,syn}} = 0.35\,\text{bar}$ gauge, $a_{\mathrm{syn}} = 0.8\,\text{kPa}^{-1}$). 
        Simulation code and configuration files are available online\footnote{\url{https://github.com/RobinWrob/On-Energy-Based-Modelling-for-Fluidic-Excitable-Neurons/releases/tag/v0.2}}.

    \subsection{Modulated restrictor}

\begin{figure}[htbp]
    \centering
    \begin{tikzpicture}[
        scale=1,
        valve body/.style={draw=black!70, fill=gray!10, rounded corners=2pt, thick},
        membrane/.style={draw=blue!80, thick, decorate, decoration={snake, amplitude=0.5mm, segment length=5mm}},
        slider/.style={draw=black!80, fill=gray!60, thick, rounded corners=1pt},
        flow arrow/.style={-{Stealth[scale=1.2]}, thick, purple!80},
        ctrl arrow/.style={-{Stealth[scale=1.5]}, very thick, blue!60},
        cross out/.style={draw=red!80, ultra thick}
    ]

        \draw[valve body] (0, 0) rectangle (4, 2.5);
        
        \draw[thick, black!70] (0, 1.5) -- (4, 1.5);
        \draw[thick, black!70] (1.0, 0) -- (1.0, .8);
        \draw[thick, black!70] (3.0, 0) -- (3.0, .8);
        \draw[thick, black!70] (1.2, 0) -- (1.2, .6);
        \draw[thick, black!70] (2.8, 0) -- (2.8, .6);
        \draw[thick, black!70] (1, .8) -- (3, .8);
        \draw[thick, black!70] (1.2, .6) -- (2.8, .6);

        \draw[membrane] (0.8, 2) -- (1.8, 2);
        \draw[membrane] (2.2, 2.0) -- (3.2, 2.0);
        \draw[thick, black!70] (0, 2) -- (0.8, 2);
        \draw[thick, black!70] (3.2, 2) -- (4, 2);
        
        \draw[slider] 
            (1.8, 0.5) -- (1.8, 1.0) --       
            (1.9, 1.2) -- (1.9, 1.4) --         
            (1.8, 1.7) -- (1.8, 2.0) --       
            (2.2, 2.0) -- (2.2, 1.7) --       
            (2.1, 1.4) -- (2.1, 1.2) --         
            (2.2, 1.0) -- (2.2, 0.5) -- cycle; 
        
        \draw[ctrl arrow] (2, 2.9) -- (2, 2.1) node[midway, right, pos=.2, xshift=2mm, font=\large\bfseries, text=black] {$P_{\mathrm{ctrl}}$};
        
        \draw[flow arrow] (1.1, -0.3) -- (1.1, 0.7) -- (1.65, 0.7);
        
        \draw[flow arrow, dashed, opacity=0.5] (2.35, 0.7) -- (2.9, 0.7) -- (2.9, -0.3) node[midway, below, pos=1, xshift=2mm, font=\large\bfseries, text=black, opacity=1] {$Q_{\mathrm{x}}$};
        

    \end{tikzpicture}
    \caption{Pneumatic realisation of the memductive elements ($\text{MR}$). The control pressure $P_{\mathrm{ctrl}}$ (driven by $\emem$ or $\es$) modulates the position of the slider through deflection of the diaphragm (blue). Flow $Q_x$ is blocked by the slider profile and permitted when the slider aligns its narrow region with the flow channel. The resulting nonlinear conductance profile is jointly determined by the slider geometry and the pressure--deflection characteristic of the diaphragm.}
    \label{fig:valve_schematic}
\end{figure}
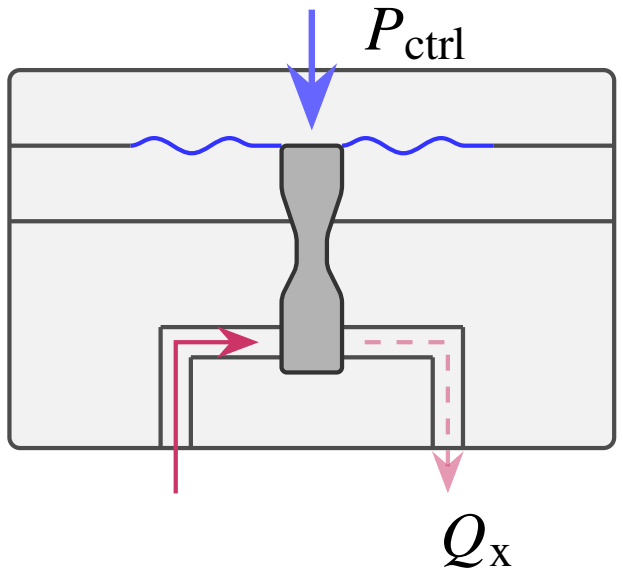

        We implement the gated branches as custom three-terminal \texttt{ModulatedRestrictor} components, with two flow terminals at pressures $p_0$ and $p_1$ and an unloaded control terminal at pressure $p_c$.
        A possible physical realisation is shown in Fig.~\ref{fig:valve_schematic}. 
        The control pressure deflects the elastic membrane which moves a slider across the flow channel.
        The slider profile regulates the channel between fully closed and open states, which, together with the pressure--deflection characteristic of the diaphragm, defines the shape and operating point of the gating nonlinearity.

        The \texttt{ModulatedRestrictor} component is defined as follows:
        the control pressure sets a dimensionless opening factor
        \begin{equation}
            \alpha(p_c)
            =\alpha_{\max}
            \sigma\bigl(\kappa(p_c-p^{\mathrm{th}})\bigr),
            \label{eq:restrictor_gate}
        \end{equation}
        while the component obeys the quasi-steady constitutive law
        \begin{equation}
            \dot{m}_0
            =\frac{\bar{\rho}}{R_0}\,
              \alpha(p_c)(p_0-p_1),
            \qquad
            R_0=\frac{128\mu L}{\pi d^4},
            \label{eq:modulated_restrictor}
        \end{equation}
        together with mass conservation $\dot{m}_0+\dot{m}_1=0$ and zero control flow $\dot{m}_c=0$.
        Here, $\bar{\rho}$ is evaluated at the mean absolute terminal pressure using the isothermal ideal-gas relation, while $R_0$ serves as a nominal first-order scaling law parameterised by the effective lumped length $L$, diameter $d$, and dynamic viscosity $\mu$.
        
        Thus, the control signal modulates the effective conductance $\alpha(p_c)/R_0$ without introducing an additional dynamical state.
        In the simulated network, the excitatory restrictor connects the membrane chamber to the excitatory supply and is controlled by $\emem$.
        The inhibitory restrictor connects the membrane chamber to the inhibitory supply and is controlled by $\es$.
        These components represent idealised pressure-controlled valves, with control-port loading neglected.

    \subsection{Spiking neuron}
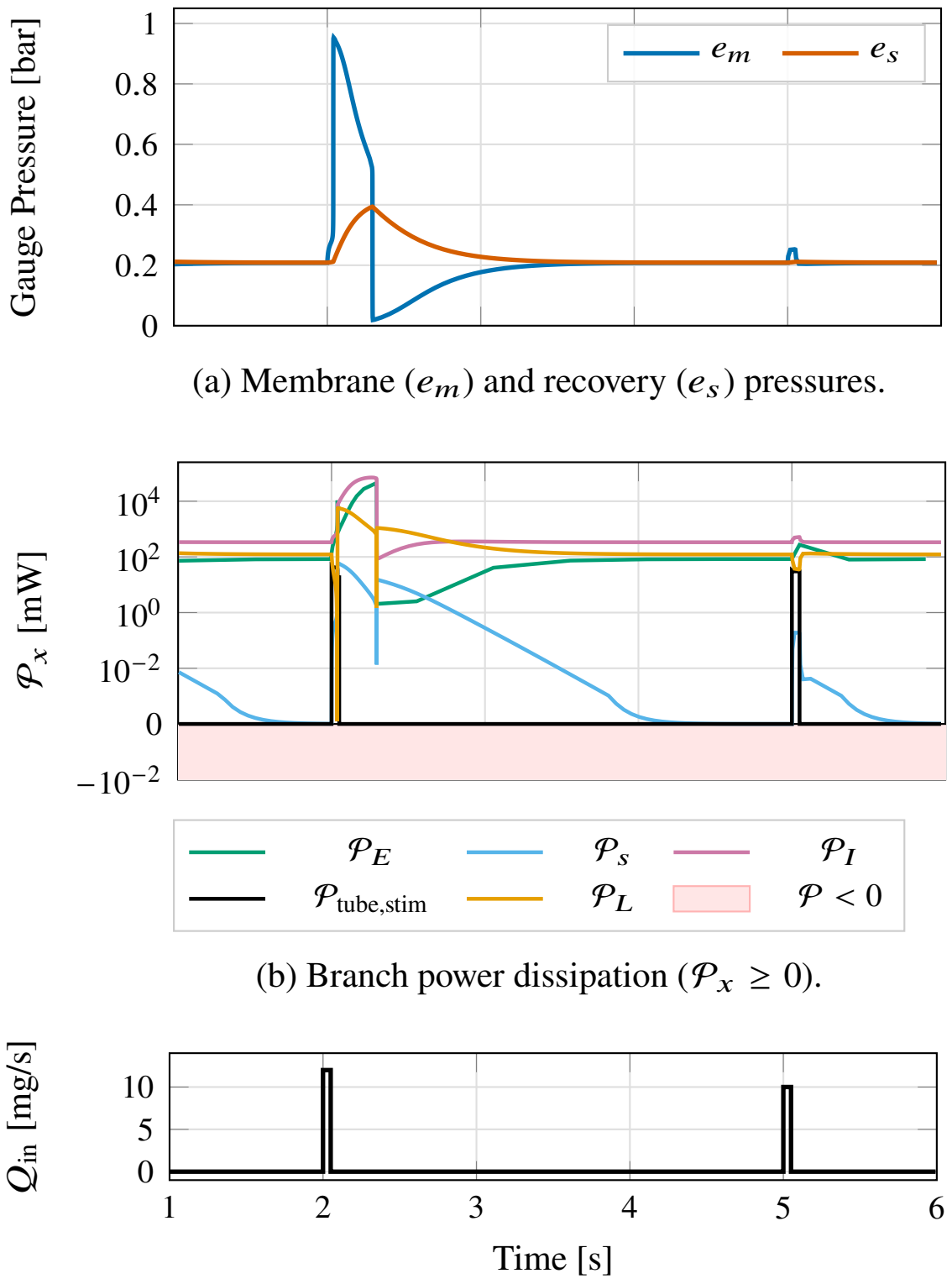
\begin{figure}[t]
    \centering
    \def\csvpath{./ECC_fig4.csv}

    \begin{subfigure}{\columnwidth}
        \centering
        \hspace{-1cm}
        \begin{tikzpicture}
            \begin{axis}[
                width=0.7\columnwidth,
                height=2.5cm,
                scale only axis=true,
                xmin=1, xmax=6,
                ymin=0.0, ymax=1.05,
                ylabel={Gauge Pressure [bar]},
                ylabel shift=-8pt,
                xtick={1,2,3,4,5,6,7,8},
                xticklabels={}, 
                ytick={0.0, 0.2, 0.4, 0.6, 0.8, 1.0},
                tick label style={font=\scriptsize},
                yticklabel style={text width=0.95cm, align=right},
                label style={font=\footnotesize},
                grid=major,
                grid style={gray!25, thin},
                legend style={
                    font=\scriptsize,
                    at={(0.98,0.95)},
                    anchor=north east,
                    legend columns=2,
                    draw=gray!40,
                    fill=white,
                    fill opacity=0.85,
                    text opacity=1,
                    /tikz/every even column/.append style={column sep=4pt}
                },
            ]

                \addplot[color=mplC0, line width=1.0pt, each nth point=5] table[
                    col sep=comma,
                    x=time_s,
                    y expr=\thisrow{e_m_gauge_Pa}*1e-5
                ] {\csvpath};
                \addlegendentry{$e_m$}

                \addplot[color=mplC1, line width=1.0pt, each nth point=5] table[
                    col sep=comma,
                    x=time_s,
                    y expr=\thisrow{e_s_gauge_Pa}*1e-5
                ] {\csvpath};
                \addlegendentry{$e_s$}
            \end{axis}
        \end{tikzpicture}
        \vspace{-0.15cm}
        \caption{Membrane ($e_m$) and recovery ($e_s$) pressures.}
        \label{fig:spike_pressures}
    \end{subfigure}

    \vspace{0.2cm} 

    \pgfmathdeclarefunction{symlog}{1}{%
        \pgfmathparse{
            (#1 >= 1e-3) ? (1 + ln(#1/1e-3)/ln(10)) : (
                (#1 <= -1e-3) ? (-1 - ln(-#1/1e-3)/ln(10)) : (#1/1e-3)
            )
        }%
    }

    \begin{subfigure}{\columnwidth}
        \centering
        \vspace{0.25cm}
        \hspace{-1cm}
        \begin{tikzpicture}
            \begin{axis}[
                width=0.7\columnwidth,
                height=2.5cm,
                scale only axis=true,
                xmin=1, xmax=6,
                ymin=-2.0, ymax=9.4,
                ylabel={$\mathcal{P}_x $ [mW]},
                ylabel shift=-8pt,
                xtick={1,2,3,4,5,6,7,8},
                xticklabels={}, 
                ytick={-2, 0, 2, 4, 6, 8},
                yticklabels={
                    $-10^{-2}$, $0$,
                    $10^{-2}$, $10^{0}$, $10^{2}$, $10^{4}$
                },
                tick label style={font=\scriptsize},
                yticklabel style={text width=0.95cm, align=right},
                label style={font=\footnotesize},
                grid=major,
                grid style={gray!25, thin},
                legend to name=legpower,
                legend columns=3,
                legend style={
                    font=\scriptsize,
                    draw=gray!40,
                    fill=white,
                    /tikz/every even column/.append style={column sep=6pt},
                    /tikz/column sep=8pt
                },
            ]
                \fill[red!10] (axis cs:1.0, -2.0) rectangle (axis cs:8.0, 0.0);

                \draw[black, thin] (axis cs:1.0, 0.0) -- (axis cs:8.0, 0.0);

                \addplot[color=mplC4, line width=0.85pt, each nth point=50] table[
                    col sep=comma,
                    x=time_s,
                    y expr={symlog(\thisrow{P_E_W}*1e3)}
                ] {\csvpath};
                \addlegendentry{$\mathcal{P}_E$}

                \addplot[color=mplC7, line width=0.85pt, each nth point=5] table[
                    col sep=comma,
                    x=time_s,
                    y expr={symlog(\thisrow{P_s_W}*1e3)}
                ] {\csvpath};
                \addlegendentry{$\mathcal{P}_s$}

                \addplot[color=mplC5, line width=0.85pt, each nth point=5] table[
                    col sep=comma,
                    x=time_s,
                    y expr={symlog(\thisrow{P_I_W}*1e3)}
                ] {\csvpath};
                \addlegendentry{$\mathcal{P}_I$}

                \addplot[color=mplC8, line width=0.85pt, each nth point=5] table[
                    col sep=comma,
                    x=time_s,
                    y expr={symlog(\thisrow{P_stim_W}*1e3)}
                ] {\csvpath};
                \addlegendentry{$\mathcal{P}_{\text{tube,stim}}$}

                \addplot[color=mplC6, line width=0.85pt, each nth point=5] table[
                    col sep=comma,
                    x=time_s,
                    y expr={symlog(\thisrow{P_L_W}*1e3)}
                ] {\csvpath};
                \addlegendentry{$\mathcal{P}_L$}

                \addlegendimage{area legend, fill=red!10, draw=red!30}
                \addlegendentry{$\mathcal{P}<0$}
            \end{axis}
        \end{tikzpicture}\\[2pt]
        \pgfplotslegendfromname{legpower}
        \caption{Branch power dissipation ($\mathcal{P}_x \ge 0$).}
        \label{fig:spike_dissipation}
    \end{subfigure}

    \vspace{0.2cm} 

    \begin{subfigure}{\columnwidth}
        \centering
        \hspace{-1cm}
        \begin{tikzpicture}
            \begin{axis}[
                width=0.7\columnwidth,
                height=1cm, 
                scale only axis=true,
                xmin=1, xmax=6,
                ymin=-1.0, ymax=14.0,
                xlabel={Time [s]}, 
                ylabel={$Q_{\mathrm{in}}$ [mg/s]},
                ylabel shift=-8pt,
                xtick={1,2,3,4,5,6,7,8},
                ytick={0, 5, 10},
                tick label style={font=\scriptsize},
                yticklabel style={text width=0.95cm, align=right},
                label style={font=\footnotesize},
                xlabel style={yshift=1pt},
                grid=major,
                grid style={gray!25, thin},
                legend style={
                    font=\scriptsize,
                    at={(0.98,0.95)},
                    anchor=north east,
                    draw=gray!40,
                    fill=white,
                    fill opacity=0.85,
                    text opacity=1
                },
            ]

                \addplot[ color=black, line width=1pt, const plot ] table[ col sep=comma, x=time_s, y expr=\thisrow{Q_in_kg_s}*1e6 ] {\csvpath};
                
            \end{axis}
        \end{tikzpicture}
        \vspace{-0.15cm}
        \caption{Input stimulus flow pulses.}
        \label{fig:spike_input}
    \end{subfigure}
    
    \caption{Fluidic EI-neuron response to supra-threshold ($t=2\,\text{s}$, $12\,\text{mg/s}$) and sub-threshold ($t=5\,\text{s}$, $10\,\text{mg/s}$) $50\,\text{ms}$ flow pulses. (a)~Only the supra-threshold pulse elicits a pressure spike. (b)~All branch powers remain non-negative. (c)~The input flow stimulus.}
    \label{fig:spike}
\end{figure}

        After interconnecting the elements following the bond graph topology, we can simulate the fluidic node's response to transient input flow pulses $\fin(t)$.
        As shown in Fig.~\ref{fig:spike_pressures}, a supra-threshold input triggers an all-or-nothing pressure spike.
        The resulting trajectory exhibits the classic mixed-feedback phases: a fast pressurisation of the membrane chamber ($\emem$) via the excitatory restrictor, followed by delayed venting as the slow recovery chamber ($\es$) pressurises and opens the inhibitory restrictor. 
        Subthreshold inputs do not trigger a spike, and the membrane pressure simply relaxes back to its resting value.
        
        Because the network is derived from physical storage and dissipation elements, the model is energetically consistent.
        Computing the instantaneous power flows during the spike confirms that dissipation across the memductive branches remains strictly non-negative ($\mathcal{P}_x = \Delta P_x Q_x \ge 0$) at all times, ensuring the action potential can be physically realised without artificial state resets.

        Unlike purely phenomenological spiking models, the bond graph representation explicitly resolves the instantaneous power dissipation across each branch, providing a direct metric to quantify pneumatic energy consumption per spike~\cite{gawthropBondGraphModelling2017a,rossControllingSimulatingSoft2016}.
        
    \subsection{Synaptic connections}
        In order to form networks, we introduce active synapses (corresponding to biological chemical synapses), modelled as additional memductive branches connected to the postsynaptic membrane node ($C_{\mathrm{m,post}}$).
        Implementation-wise, a synapse is simply another \texttt{ModulatedRestrictor} placed in series with a constant reversal pressure source ($Se{:}\Esyn$). 
        However, unlike intrinsic channels, the conductance of the synaptic valve $\gsyn$ is modulated by a \textit{presynaptic} state (e.g., $p_c = e_{\mathrm{s,pre}}$).
        This coupling is shown in the semi-word bond graph of Figure~\ref{fig:hco_bondgraph}, where the neuron bond graph is represented by a `neuron' element.
        The control connection is idealised as an unloaded information signal, without direct mass transfer between the internal chambers of the neurons.
        Opening a synaptic conductance drives the postsynaptic pressure towards the reversal pressure $\Esyn$, which may be selected to provide excitation ($\EE \geq \Esyn>\EL$) or inhibition ($\EI\leq\Esyn<\EL$).

    \subsection{Half-Center oscillator}
        To demonstrate how these elements can be used to form a CPG, we couple two identical fluidic EI neurons symmetrically via mutual inhibitory synapses. 
        Rhythm generation relies on post-inhibitory rebound~\cite{burgerExcitabilityScalesUnifying2026}: when one neuron fires, its synaptic conductance inhibits the other.
        During inhibition, the recovery effort decreases, reducing activation of the intrinsic inhibitory conductance.
        Upon release, recovery of the membrane pressure allows the fast excitation branch to trigger a rebound spike.
        Together, mutual inhibition and post-inhibitory rebound produce alternating activity.
        
\begin{figure}[t]
    \centering
    
    \begin{subfigure}{\columnwidth}
        \centering
        \begin{tikzpicture}[
            >=latex,
            el/.style={draw, rectangle, line width=0.5pt, minimum width=1.0cm, minimum height=0.55cm, inner sep=2pt, font=\small},
            word/.style={draw, rectangle, line width=0.8pt, minimum width=2.2cm, minimum height=0.8cm, inner sep=4pt, font=\small\bfseries, rounded corners=2pt, fill=black!5},
            jn/.style={inner sep=2pt, minimum size=0.5cm, font=\small\bfseries},
            lbl/.style={font=\scriptsize, inner sep=1.5pt},
            bondCH/.style={line width=0.6pt, -{Straight Barb[length=2mm, width=1.6mm, left] Bar[width=3mm, line width=0.6pt]}},
            bondCT/.style={line width=0.6pt, {Bar[width=3mm, line width=0.6pt]}-{Straight Barb[length=2mm, width=1.6mm, left]}},
            sig/.style={line width=0.6pt, -{Latex[length=1.6mm]}, rounded corners=1.5mm}
        ]
            \definecolor{sigGreen}{RGB}{0,122,31}
            \definecolor{sigRed}{RGB}{213,11,11}

            \node[word] (NA) at (0, 0.7) {Neuron A};
            \node[word] (NB) at (6, 0.7) {Neuron B};

            \node[jn] (JA)  at (0, 2.2) {1};
            \node[el] (SeA) at (0, 3.4) {$\mathrm{Se}\!:\!E_{\mathrm{syn}}$};
            \node[el] (MRA) at (2, 2.2) {$\mathrm{MR}\!:\!R_{\mathrm{syn,A}}$};

            \node[jn] (JB)  at (6, 2.2) {1};
            \node[el] (SeB) at (6, 3.4) {$\mathrm{Se}\!:\!E_{\mathrm{syn}}$};
            \node[el] (MRB) at (4, 2.2) {$\mathrm{MR}\!:\!R_{\mathrm{syn,B}}$};

            \draw[bondCH] (SeA) -- node[lbl, left, pos=0.4] {$E_{\mathrm{syn}}$} node[lbl, right, pos=0.4] {$f_{\mathrm{syn,A}}$} (JA);
            \draw[bondCT] (JA)  -- node[lbl, left] {$e_{m,A}$} node[lbl, right] {$f_{\mathrm{syn,A}}$} (NA);
            \draw[bondCH, shorten >=2pt] (JA) -- node[lbl, above, pos=0.4] {$e_{R,A}$} node[lbl, below, pos=0.4] {$f_{\mathrm{syn,A}}$} (MRA);
       
            \draw[bondCH] (SeB) -- node[lbl, right, pos=0.4] {$E_{\mathrm{syn}}$} node[lbl, left, pos=0.4] {$f_{\mathrm{syn,B}}$} (JB);
            \draw[bondCT] (JB)  -- node[lbl, right] {$e_{m,B}$} node[lbl, left] {$f_{\mathrm{syn,B}}$} (NB);
            \draw[bondCH, shorten >=2pt] (JB)  -- node[lbl, above, pos=0.4] {$e_{R,B}$} node[lbl, below, pos=0.4] {$f_{\mathrm{syn,B}}$} (MRB);

            \draw[sig, sigGreen] (NA.east) 
                -- node[lbl, above] {$e_{s,A}$} (NA.east -| MRA.south) 
                -- (MRB.south);

            \draw[sig, sigGreen] (NB.west) 
                -- node[lbl, above] {$e_{s,B}$} (NB.west -| MRB.south) 
                -- (MRA.south);
                
        \end{tikzpicture}
        \caption{Bond graph interconnection of two neurons with mutual inhibition using modulated resistors.}
        \label{fig:hco_bondgraph}
    \end{subfigure}
    
    \vspace{0.4cm} 

    \begin{subfigure}{\columnwidth}
        \centering
        \hspace{-20pt}
        \def\csvpath{./ECC_fig5b.csv}
        \begin{tikzpicture}
            \begin{axis}[
                width=0.75\columnwidth,
                height=3.0cm,
                scale only axis=true,
                xmin=0, 
                xmax=10,
                xlabel={Time [s]},
                ylabel={Gauge Pressure [bar]},
                ylabel shift=-16pt,
                tick label style={font=\scriptsize},
                yticklabel style={text width=0.95cm, align=right},
                label style={font=\footnotesize},
                grid=major,
                grid style={gray!25, thin},
                legend style={
                  font=\scriptsize,
                  at={(0.98,0.95)},
                  anchor=north east,
                  draw=gray!40,
                  fill=white,
                  fill opacity=0.85,
                  text opacity=1,
                },
              ]
              
              \addplot[
                color=mplC0, 
                line width=1.0pt,
                each nth point=10
              ] table[
                col sep=comma, 
                x=time_s, 
                y expr=\thisrow{em_A_gauge_Pa}*1e-5
              ] {\csvpath};
              \addlegendentry{$e_{m,A}$}
              
              \addplot[
                color=mplC1, 
                line width=1.0pt,
                each nth point=10
              ] table[
                col sep=comma, 
                x=time_s, 
                y expr=\thisrow{em_B_gauge_Pa}*1e-5
              ] {\csvpath};
              \addlegendentry{$e_{m,B}$}
              
            \end{axis}
        \end{tikzpicture}
        \caption{Membrane-pressure traces.}
        \label{fig:hco_traces}
    \end{subfigure}
    
    \caption{Bond graph design of a half-centre oscillator (HCO). (a) Bond graph representation of the interconnection, showing mutual inhibition via modulation signals $e_{s,A}$ and $e_{s,B}$ (green). (b) Membrane-pressure traces of the two neurons.}
    \label{fig:hco}
\end{figure}
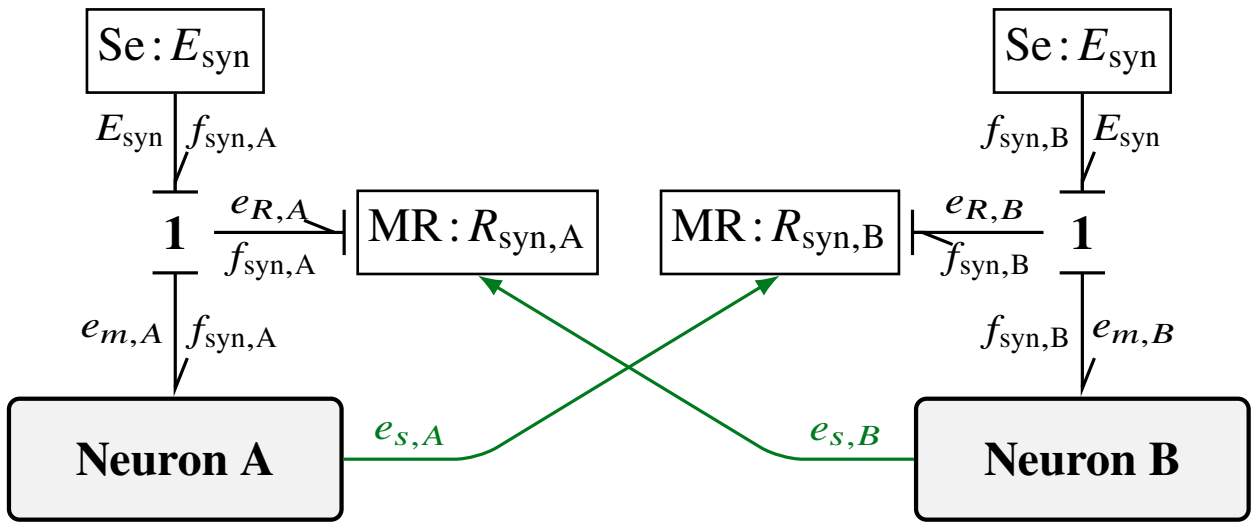
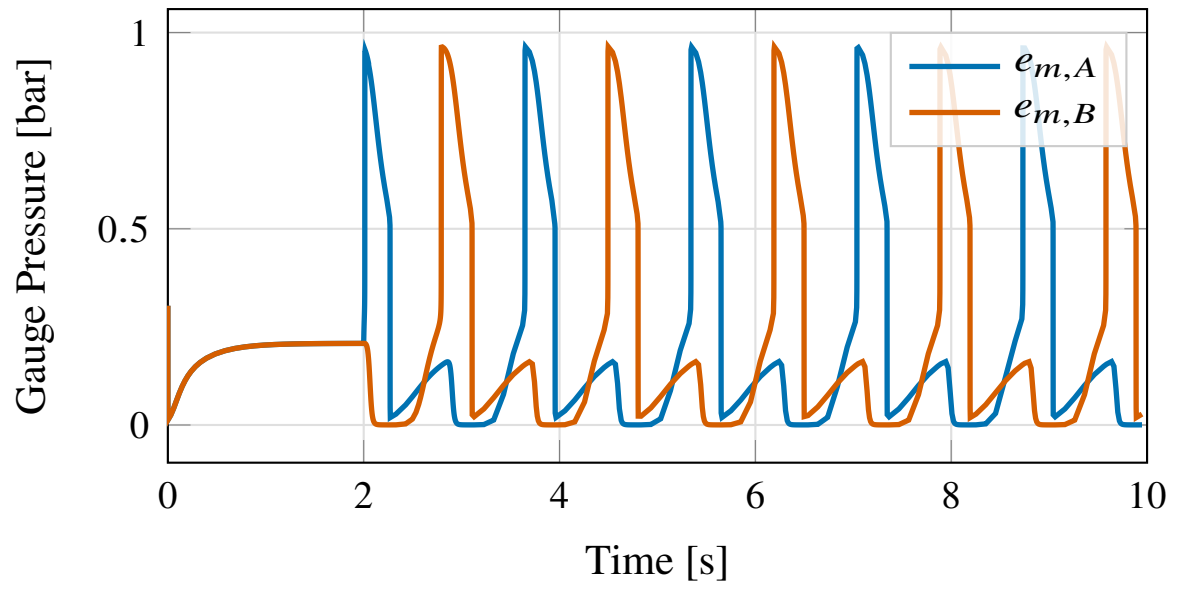

        As shown in Fig.~\ref{fig:hco_traces}, a brief asymmetric startup perturbation initiates sustained alternating pressure spikes ($e_{\mathrm{m,A}}$ versus $e_{\mathrm{m,B}}$) at timescales directly applicable to embodied soft robotic locomotion.
        The resulting half-center oscillator illustrates how the proposed excitable nodes can be interconnected to generate rhythmic control signals.

\section{Conclusion and future work}
    We presented an energy-based framework to translate neuronal excitability into continuous fluidic media at behavioural timescales. 
    By shifting away from abstract integrate-and-fire models and grounding ourselves in the Hodgkin--Huxley formulation through bond graph theory, we established a physically consistent blueprint comprising explicit energy storage, supply, and state-dependent dissipation elements. 
    Through lumped pneumatic simulations, we validated that mapping these elements to fluidic chambers and pressure-controlled valves yields energetically consistent action potentials and post-inhibitory rebound dynamics without relying on unphysical state resets.
    Future work will focus on physically implementing such event-based fluidic networks, specifically aiming to synthesize closed-loop central pattern generators for reflexive and adaptive soft robotic locomotion.

\printbibliography

\end{document}